# THE LUMINOSITIES OF SUPERNOVAE OF TYPE Ia

# DERIVED FROM CEPHEID CALIBRATIONS

Sidney van den Bergh

Dominion Astrophysical Observatory

5071 West Saanich Road

Victoria, British Columbia, V8X 4M6, Canada

Electronic mail:  vandenbergh@dao.nrc.ca

# ABSTRACT

Available data on the luminosities of supernovae of Type Ia (SNe Ia) that have been calibrated by Cepheids are collected and discussed.  The objects in the present sample show a range of ~20 in luminosity.  The data strongly confirm the suspicion that SNe Ia in early-type systems are, on average, fainter than those that occur in late-type galaxies.  Historical observations of S Andromedae suggest that the maximum magnitude versus rate-of-decline relationship for SNe Ia  exhibits a large intrinsic dispersion.  This is not surprising because the amount of $^{56}$Ni that is produced, and other observable properties of SNe Ia, are expected to depend sensitively on how much fuel is ignited.

Subject headings:  cosmology-distance scale - stars:  supernovae:  general - stars: variables:  Cepheids

# 1. INTRODUCTION

The first great distance scale controversy (Curtis 1921, Shapley 1921) was resolved when Edwin Hubble (1925) discovered Cepheid variables in M31 and M33. Ever since that time Cepheids have remained the cornerstone to the determination of extragalactic distances (Sandage 1972). Can Cepheids resolve the second great distance scale controversy between astronomers who believe that $H_o \approx 55$ km s$^{-1}$ Mpc$^{-1}$ and those who think that $H_o \approx 80$ km s$^{-1}$ Mpc$^{-1}$? Even with the Hubble Space Telescope (HST), it is not possible to study Cepheids out to distances at which the random motions of galaxies are negligible compared to the cosmic Hubble flow. However, it might be possible to use Cepheids to calibrate other more luminous distance indicators that can be observed out to cosmologically significant distances. In the present paper, an attempt is made to marshall all presently available data on extragalactic Cepheids to calibrate the luminosities of SNe Ia.

# 2. DISTANCES TO INDIVIDUAL SUPERNOVAE

## 2.1 SN1981B in NGC 4536

The well-observed normal Type Ia supernovae 1981B is presently one of the most suitable objects for the determination of the luminosities of SNe Ia, and hence for estimation of the extragalactic distance scale. In a recent discussion of the light curve of this object Schaefer (1995b) obtained B(max) = 12.04 ± 0.04,

(B-V)$^{max}$ = 0.04 ± 0.06 and $\Delta m_{15}$ = 1.07 ± 0.09 mag (were $\Delta m_{15}$ is the decline in B

magnitude during the 15 days following maximum). Vacca & Leibundgut (1995)

have re-discussed the light curve of SN1981B and find B(max) = 12.05, (B-V)$^{max}$ =

+0.09 and $\Delta m_{15}$ = 1.14, which is in reasonable agreement with Schaefer's results.

However, very recently Patat et al. (1996a) obtained B(max) = 11.74 and (B-V)$^{max}$

= +0.06 ± 0.05 by recalibrating the comparison stars used to derive the light curve

of SN1981B. Turatto et al. (1996) give a value $\Delta m_{15}$ = 1.10 for this object, which

agrees well with the two values cited above. HST observations of the Cepheids in

NGC 4536 by Saha et al. (1996) yield a distance modulus $(m\text{-}M)_o$ = 31.10 ± 0.13

for the parent galaxy of SN1981B. Assuming $(B\text{-}V)_o^{max}$ = 0.00 [see Patat et al.

(1996a) and references therein] and A$_B$ = 4.1 E$_{B\text{-}V}$ (Mathis 1990) the data by Patat

et al. yield M$_B$ (max) = -19.61 ± 0.24. The data on SN1981B obtained above, and

those for additional supernovae which will be discussed below, are summarized in

Table 1.

**2.2**    <u>SN1960F in NGC 4496</u>

Another supernova in a galaxy for which a Cepheid distance has recently

become available is SN1960F in NGC 4496. According to Saha et al. (Saha

1995), the Cepheid distance modulus of this object is $(m\text{-}M)_o$ = 31.1 ± 0.15.

Unfortunately, the light curve of SN1960F (Leibundgut et al. 1991) is not very well constrained by observations which indicate that B(max) ≈ 11.7. From a recent re-discussion of the light curve of SN1960F Schaefer (1996c) concludes that B(max) = 11.77 ± 0.07, (B-V)$^{max}$ = 0.26 ± 0.19 and that $\Delta m_{15}$ = 1.06 ± 0.08. Assuming $(B-V)^{max}_{o}$ = 0.0 these values yield $M_B$(max) = -20.40 ± 0.79, in which the uncertainty is mainly due to the large mean error in the observed color of SN1960F at maximum light. The value of $M_B$(max) derived from these data is listed in Table 1. NGC 4496 is located in the "southern extension" of the Virgo cluster.

**2.3**    <u>SN1885 in M 31</u>

It is sometimes forgotten (van den Bergh 1994) that S Andromedae = SN1885 belongs to the small band of SNe for which Cepheid-based distances are available. In their definitive discussion de Vaucouleurs & Corwin (1985) suggested that S And was probably a supernova of Type Ia. They write "nearly all the features noted, however marginally, in the spectrum of S And by spectroscopists of 1885 correspond closely to known intensity maxima or minima in the spectra of Type I supernovae". De Vaucouleurs & Corwin also write that "It is difficult to say whether the λ 6150 absorption [which is diagnostic of SNe Ia], if present, would have been missed by the spectroscopists of 1885. Nearly all were on the lookout for emission lines; only Vogel reported the two minima near 4500 Å and 5700 Å.  Absence of evidence is not evidence of absence."  In any case, the fact that S And occurred in a region of the bulge of M31 which is devoid of star formation militates against it having been an SN Ibc or SN II, which are believed to have massive progenitors.  De Vaucouleurs & Corwin derive V(max) = 5.85 for S And.  From the data plotted in their Fig. 5, I <u>estimate</u> that the uncertainty in this value is ~ 0.3 mag.  After correcting for a foreground reddening $E_{B-V} = 0.08$ (van den Bergh 1991), and assuming no additional extinction in the nuclear bulge of M31, one obtains $V_o(max) = 5.6 \pm 0.3$.  In conjunction with a distance modulus $(m\text{-}M)_o = 24.3 \pm 0.1$, this yields $M_V(max) = -18.70 \pm 0.32$.  Unfortunately, no reliable color observations of S And were obtained near maximum light.  The "probable B-band light curve" of S And plotted in Fig. 9 of de Vaucouleurs & Corwin has B(max) = 7.3 and $\Delta M_{15} \approx 2.1$ mag.  With the distance modulus and reddening value cited above this gives $B_o(max) = 6.97$ and $M_B(max) = -17.33$.  The uncertainty in this value of $M_B(max)$ is difficult to estimate.  However, it is clear that S And (if it was indeed a supernova of type Ia) is among the least luminous SNe Ia listed in Table 1.

**2.4**   <u>SN1972E and SN1895B in NGC 5253</u>

From observations of Cepheid variables in NGC 5253, Saha et al. (1995)
find $(m\text{-}M)_V = 28.10 \pm 0.07$ (internal error).  Including a possible error of 0.1 mag
in the distance modulus of the LMC in quadrature increases the total uncertainty of
this distance modulus to $\pm 0.12$ mag.  For SN1972E Leibundgut et al. (1991a)
adopt V(max) = $8.60 \pm 0.1$ (the real uncertainty of this value may, however, be
somewhat larger because systematic observations of SN1972E did not start until a
number of days after maximum).  From the data given above, Saha et al. derive
$M_V$ (max) = -19.50 $\pm$ 0.21, in which the quoted uncertainty includes allowance for
a possible <u>small</u> difference between the reddening of SN1972E and the mean
reddening of the Cepheids in this galaxy.  With B(max) - V(max) =  -0.02
(Leibundgut et al. 1991a) and $E_{B\text{-}V}$ = 0.03, one obtains $M_B$ (max) = -19.55 $\pm$ 0.21.
From archival data Schaefer (1995a) finds that SN1895B = Z Centauri, which also
occurred in NGC5253, may have been a few tenths of a magnitude brighter than
SN1972E at maximum.

From trial fits to five template light curves Hamuy et al. (1995) find $\Delta m_{15}$ =
0.94 $\pm$ 0.10 mag for SN1972E.  This value is, however, quite uncertain because no
B observations are available at or near the time of maximum light.

**2.5**     SN1937C in IC 4182

From observations of 28 Cepheids, Saha et al. (1994) derive

$(m$-$M)_o = 28.36 \pm 0.05$ and $A_v = 0.0$ for the low luminosity galaxy IC 4182.

Adding a zero point uncertainty of 0.1 mag in quadrature, this yields a true

distance modulus uncertainty of $\pm 0.11$ mag.  From a re-discussion of the

observations of SN1937C, Saha et al. find B(max) = 8.83 $\pm$ 0.11 and V(max) =

8.64 $\pm$ 0.10.  However, these values remain controversial.  From recent re-measurement and re-analysis of Palomar Schmidt observations of SN1937C,

Pierce & Jacoby (1995) derive significantly fainter values B(max) = 8.94 $\pm$ 0.03

and V(max) = 9.00 $\pm$ 0.03; but see Schaefer (1996a) who finds B(max) = 8.71 $\pm$

0.14 and V(max) = 8.72 $\pm$ 0.06.  The corresponding absolute magnitudes are

$M_B$(max) =  -19.53  $\pm$  0.16 and $M_v$(max) =  -19.72  $\pm$  0.15 from the Saha et al.

apparent magnitudes, and $M_B$(max) =  -19.42  $\pm$  0.11 and $M_v$(max) = -19.36 $\pm$

0.11  from the apparent magnitudes derived by Pierce & Jacoby (1995).  Pierce &

Jacoby find $\Delta m_{15} = 1.07 \pm 0.11$, whereas Schaefer obtains $\Delta m_{15} = 0.84 \pm 0.03$.  A value $\Delta m_{15} = 0.95$: will be adopted.

**2.6**     SN1990N in NGC 4639

NGC 4639 is a spiral of type SBb that is located $\sim 4°$ east of M87 (which is generally taken to define the position of the center of the Virgo cluster).  From observations of Cepheids in NGC 4639 Sandage et al. (1996) find $(m\text{-}M)_o = 32.00$

$\pm$ 0.23. Excellent photometry of SN1990N is provided by Leibundgut et al.

(1991b). After making a small correction for Galactic foreground absorption,

Leibundgut et al. find $B_o(max) = 12.65$. Both the blue color at maximum $(B\text{-}V)^{max}$

= +0.02, and the lack of absorption lines in the spectrum of SN1990N (Jeffery,

Leibundgut & Kirshner 1992) suggest that the B(max) value of SN1990N was not

strongly affected by internal absorption. Adopting $(B\text{-}V)_o^{max} = 0.00$ yields $E_{B\text{-}V} =$

0.02 and hence $B_o(max) = 12.57$ and $M_B(max) = -19.43 \pm 0.23$. Phillips (1993)

finds $\Delta m_{15} = 1.01 \pm 0.10$ mag for SN1990N.

Yasuda, Fukugita & Okamura (1996) find that the Fisher-Tully technique

gives a distance of 22.29 $\pm$ 2.25 Mpc for NGC 4639. This places NGC 4639 well

behind the core of the Virgo cluster for which Yasuda et al. obtain a distance of 15

- 18 Mpc.

**2.7**    Supernovae in the Virgo cluster

Distances are now available (van den Bergh 1995) for Cepheids in four

spiral galaxies located in the Virgo region. These galaxies are found to have

$< (m\text{-}M)_o > = 31.02$, with an rms dispersion of only 0.08 mag. After including the

uncertainty of the distance to the Large Magellanic Cloud and possible errors in the zero-point calibration of HST in quadrature, one obtains an uncertainty of $\pm$ 0.2 mag for the distance modulus of spiral galaxies in the Virgo region, i.e. $< (m\text{-}M)_o > = 31.02 \pm 0.2$ for Virgo spirals. This value is consistent with $(m\text{-}M)_o = 31.12 \pm 0.25$ that Whitmore et al. (1995) have very recently obtained from observations of globular clusters associated with the giant elliptical galaxy M87. This result shows that there is presently no compelling evidence to suggest that Virgo elliptical and spiral galaxies are located at systematically different distances. For a more detailed discussion of the depth of the spiral galaxy distribution in the Virgo region along the line of sight the reader is referred to Yasuda, Fukugita & Okamura (1996).

Fukugita & Hogan (1991) have re-discussed the supernovae of Type Ia in the Virgo cluster that were listed by Capaccioli et al. (1990). Their compilation includes eight SNe Ia in E and S0 galaxies, one in an S(B)a and two in spirals of types Sbc - Sc (one of these is NGC 1981B, which was discussed in § 2.1). For their sample of eight SNe Ia in E and S0 galaxies (which excludes the peculiar sub-luminous object SN1991bg), one finds $< B(max) > = 12.26 \pm 0.13$, with a dispersion of 0.37 mag[1]. Adopting the Virgo distance modulus

---

---

$(m\text{-}M)_o = 31.02 \pm 0.2$, that is derived from four Cepheids in the Virgo region (van den Bergh 1995), the data by Fukugita & Hogan yield $< M_B(max) > = -18.76 \pm 0.24$ for the absolute magnitude of SNe Ia in E and S0 galaxies. Since most Virgo E and S0 galaxies are almost dust-free this mean value should not be much affected by interstellar absorption. Inspection of the light curves of five Virgo SNe Ia in E and S0 galaxies plotted by Leibundgut et al. (1991a) yields $< \Delta m_{15} > \approx 1.1$ mag for these objects.

Photometry of the peculiar sub-luminous supernova SN1991bg in the Virgo elliptical NGC4374 (= M84) has been published by Filippenko et al. (1992) and by Leibundgut et al. (1993). Using Filippenko's date of B(max) and Leibundgut's photometry yields $B(max) = 14.75$ and $\Delta m_{15} = 1.9$ mag. From a re-discussion of all observations of SN1991bg Turatto et al. (1996) have recently obtained $\Delta m_{15} = 1.95$ mag, which is the value used in Table 1. In conjunction with $(m\text{-}M)_o = 31.02 \pm 0.2$ for the Virgo cluster the observed value of B(max) gives $M_B(max) = -16.27 \pm 0.2$.

The data for this supernova, which was not included in the Virgo sample of Fukugita & Hogan (1991), are listed separately in Table 1.

SN1994D was a spectroscopically normal SN Ia that occurred recently near the edge of a dust lane in the S0 galaxy NGC 4526, which is a member of the Virgo cluster. Detailed photometry by Patat et al. (1996b) yields B(max) = 11.84, $E_{B-V}$ = $0.06 \pm 0.02$ and $\Delta m_{15}$= 1.26 mag for this object. With $B_o$(max) = 11.84 - 0.25 $\pm$ 0.08 and $(m-M)_o$ = $31.02 \pm 0.20$. This gives $M_B$(max) = $-19.43 \pm 0.22$. The S0 galaxy NGC 4526 probably belongs to the compact E/S0 core of the Virgo cluster, rather than to its more extended envelope of spirals. Depth along the line of sight is therefore not expected to make a major contribution to the uncertainty in the $M_B$(max) value for SN1994D.

## 3.    DISCUSSION

All presently available data on SNe Ia luminosities that have been derived from Cepheid distance calibrations are collected in Table 1. Inspection of these data shows that the $M_B$(max) values for supernovae of Type Ia exhibit a significant dispersion. A similar conclusion had previously been reached by Schaefer (1996b) who finds that "A significant fraction (probably the majority) of Type Ia events are not standard candles, with most being subluminous by 1 to 6 magnitudes". Even if the peculiar object SN1991bg and SN1960F [for which $M_B$(max) is uncertain] are excluded there still remains a range from $M_B$(max) = -19.61 for SN1981B to $M_B$(max) = -17.3 for S And.  An even larger range would be obtained if one were to include the very uncertain value $M_B$(max) = -20.4 ± 0.8 for SN1960F or $M_B$ = -19.94 ± 0.23 for SN1991T.  It is of interest to note that the two faintest SNe in Table 1 also have the largest values of $\Delta m_{15}$.  <u>The large range in $M_B$(max) and $\Delta m_{15}$ values listed in Table 1 is not unexpected because the amount of $^{56}$Ni that is produced, and other observed properties of SNe Ia, depend sensitively on how the fuel is ignited</u> (Niemeyer, Hillebrand & Woosley 1996).

The luminous objects SN1937C, SN1960F and SN1972E occurred in late-type galaxies, whereas all of the faintest SNe occurred in galaxies of early-type (or in the nuclear bulge of the spiral galaxy M31).  A similar difference in supernovae luminosities had previously been noted by Hamuy et al. (1995).  The relatively high luminosity of SN1994D (Patat et al. 1996) in the S0 galaxy NGC 4526 might be explained by the (admittedly <u>ad hoc</u>) hypothesis that its progenitor was a young object that is associated with a nearby dust lane.

I am indebted to Abi Saha for permission to quote his distance modulus to NGC 4536 in advance of publication an to Brad Schaefer for lively discussions about the distance moduli of individual supernovae.

## APPENDIX

The superluminous supernova 1991T occurred in NGC 4527. The data in Table 2 suggest that NGC 4527 might be a physical companion to NGC 4536, for which Saha et al. (1995) obtain a distance modulus $(m\text{-}M)_o = 31.05 \pm 0.15$. According to Phillips et al. (1992) this object had B(max) = $11.64 \pm 0.05$ and $(B\text{-}V)^{max} = 0.13 \pm 0.04$. With an <u>assumed</u> reddening E(B-V) = $0.13 \pm 0.04$ and $A_B / E_{B\text{-}V} = 4.1$ this yields $B_o(max) = 11.11 \pm 0.17$ and hence, if NGC 4527 and NGC 4536 are at the same distance, $M_B(max) = -19.94 \pm 0.23$. Phillips (1993) finds $\Delta m_{15} = 0.94 \pm 0.07$ for SN1991T.

TABLE 1

SNe Ia WITH CEPHEID CALIBRATIONS

| SN | Galaxy | Type | $(m\text{-}M)_o$ | $\Delta m_{15}$ | $M_B(max)$ |
|---|---|---|---|---|---|
| 1885 | M 31 | Sb | 24.3 $\pm 0.1$ | 2.1 | -17.33: |
| 1937C | IC 4182 | Ir | 28.36 $\pm 0.11$ | 0.95: | -19.53 $\pm 0.16$ |
| 1960F | NGC 4496 | SBc | 31.1 $\pm 0.15$ | 1.06 | -20.40 $\pm 0.79$ |
| 1972E | NGC 5253 | Pec | 28.10 $\pm 0.12^a$ | 0.94 | -19.55 $\pm 0.21$ |
| 1981B | NGC 4536 | Sc | 31.10 $\pm 0.13$ | 1.1 | -19.61 $\pm 0.24$ |
| ... | Virgo[b] | ... | 31.02 $\pm 0.2$ | $< 1.1 >$ | -18.76 $\pm 0.24$ |
| 1990N | NGC 4639 | Sb | 32.00 $\pm 0.23$ | 1.01 | -19.43 $\pm 0.23$ |
| [1991T | NGC 4527 | Sb | 31.05 $\pm 0.15^d$ | 0.94 | -19.99 $\pm 0.21$] |
| 1991bg[c] | NGC 4374 | E1 | 31.02 $\pm 0.2$ | 1.95 | -16.27 $\pm 0.2$ |
| 1994D | NGC 4526 | S0 | 31.02 $\pm 0.2$ | 1.26 | -19.43 $\pm 0.22$ |

[a]    Visual distance modulus.

[b]    Eight SNe in E and S0 galaxies that occurred before 1991.

[c]    Spectroscopically peculiar.

[d]    See Appendix.

TABLE 2

DATA FOR NGC 4527 AND NGC 4536[a]

| NGC | Type | $B_T$ | $\alpha$ | (1950) | $\delta$ | $V_o$(km s$^{-1}$) |
|------|-------|-------|----------|--------|----------|---------------------|
| 4527 | Sb II | 11.32 | $12^h$ $31.^m6$ | | $+ 02^o 56'$ | 1577 |
| 4536 | Sc II | 11.01 | 12  31.9 | | $+ 02$ 28 | 1646 |

[a]   Adapted from Sandage & Tammann (1981).